\DocumentMetadata{}
\documentclass[manuscript, sigconf]{acmart}

\usepackage{tabularray}
\UseTblrLibrary{booktabs}
\usepackage[english]{babel}
\usepackage{CJKutf8}
\AtBeginDocument{%
  \providecommand\BibTeX{{%
    \normalfont B\kern-0.5em{\scshape i\kern-0.25em b}\kern-0.8em\TeX}}}

\setcopyright{acmcopyright}
\copyrightyear{2018}
\acmYear{2018}
\acmDOI{XXXXXXX.XXXXXXX}

\acmBooktitle{Woodstock '18: ACM Symposium on Neural Gaze Detection,
 June 03--05, 2018, Woodstock, NY} 
\acmPrice{15.00}
\acmISBN{978-1-4503-XXXX-X/18/06}

\begin{document}

\title[Practice, Perception, and Challenge of BLV Students Learning in `Blind Colleges']{Understanding the Practice, Perception, and Challenge of Blind or Low Vision Students Learning through Accessible Technologies in Non-Inclusive `Blind Colleges'}


\author{Xiuqi Tommy Zhu}
\email{zhu.xiu@northeastern.edu}
\affiliation{%
  \institution{The Future Lab, Tsinghua University}
  \city{Beijing}
  \country{China}
}
\affiliation{%
  \institution{Northeastern University}
  \city{Boston}
  \country{United States}
}

\author{Ziyue Qiu}
\email{ziyue.qiu.23@ucl.ac.uk}
\affiliation{%
  \institution{The Future Lab, Tsinghua University}
  \city{Beijing}
  \country{China}
}
\affiliation{%
  \institution{University College London}
  \city{London}
  \country{United Kingdom}
}

\author{Ye Wei}
\email{yewei@andrew.cmu.edu}
\affiliation{%
  \institution{The Future Lab, Tsinghua University}
  \city{Beijing}
  \country{China}
}
\affiliation{%
  \institution{Carnegie Mellon University}
  \city{Pittsburgh}
  \country{United States}
}

\author{Jianhao Wang}
\email{Jwang74@tufts.edu}
\affiliation{%
  \institution{The Future Lab, Tsinghua University}
  \city{Beijing}
  \country{China}
}
\affiliation{%
  \institution{Tufts University}
  \city{Medford}
  \country{United States}
}

\author{Yang Jiao}
\email{jiaoyang7@tsinghua.edu.cn}
\authornote{Corresponding Author}
\affiliation{%
  \institution{The Future Lab, Tsinghua University}
  \city{Beijing}
  \country{China}
}


\begin{abstract}
  In developing and underdeveloped regions, many `Blind Colleges' exclusively enroll individuals with Blindness or Vision Impairment (BLV) for higher education. While advancements in accessible technologies have facilitated BLV student integration into `Integrated Colleges,' their implementation in `Blind Colleges' remains uneven due to complex economic, social, and policy challenges. This study investigates the practices, perceptions, and challenges of BLV students using accessible technologies in a Chinese `Blind College' through a two-part empirical approach. Our findings demonstrate that tactile and digital technologies enhance access to education but face significant integration barriers. We emphasize the critical role of early education in addressing capability gaps, BLV students' aspirations for more inclusive educational environments, and the systemic obstacles within existing frameworks. We advocate for leveraging accessible technologies to transition `Blind Colleges' into `Integrated Colleges,' offering actionable insights for policymakers, designers, and educators. Finally, we outline future research directions on accessible technology innovation and its implications for BLV education in resource-constrained settings.


\end{abstract}



\begin{teaserfigure}
\includegraphics[width=\textwidth]{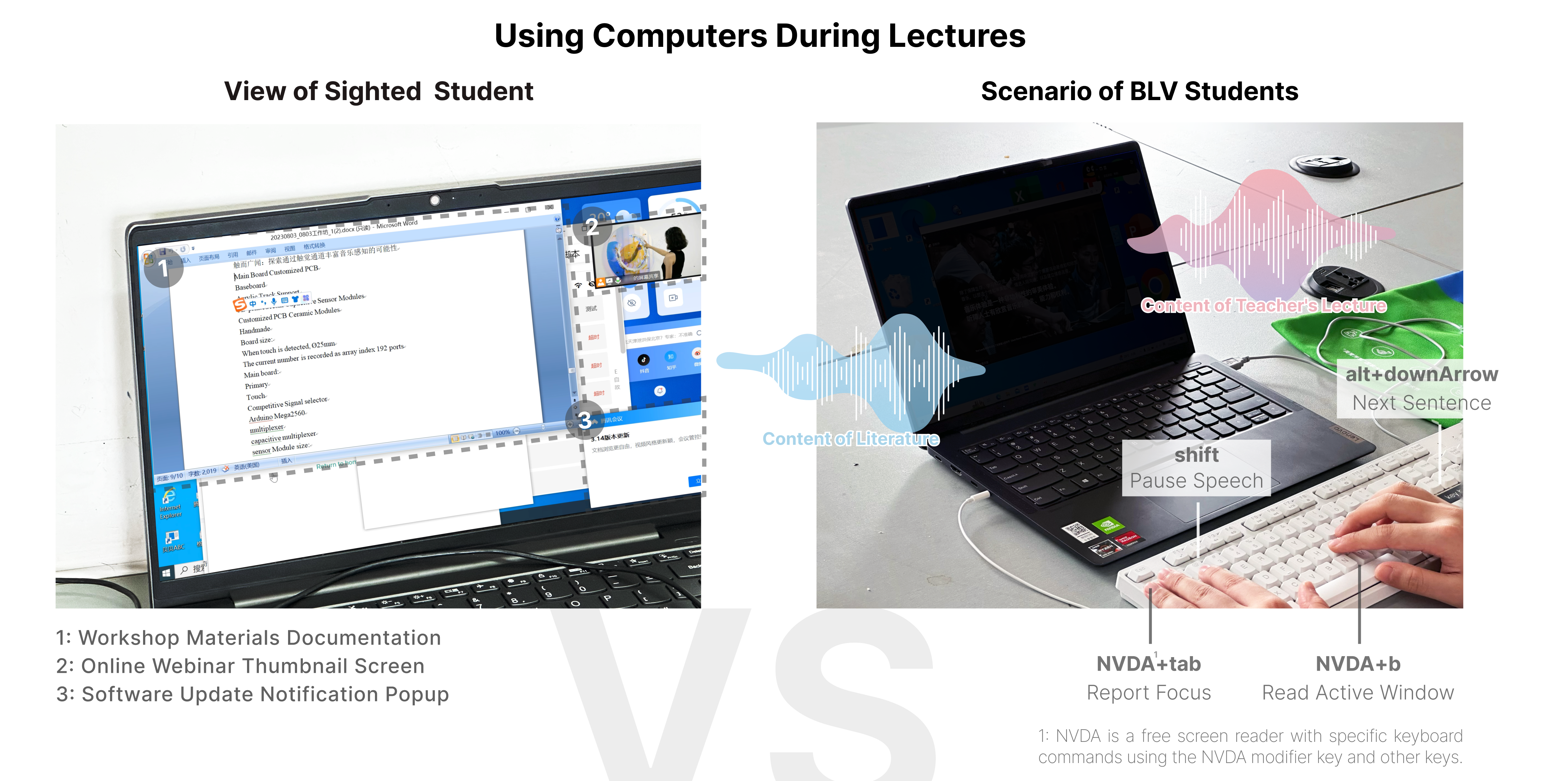}
  \caption{\textbf{This figure compares the differences in the scenarios of sighted individuals and BLV students using computers for on-line lecture learning, showcasing their distinct experiences. Individuals with sight can visually perceive multiple tasks occurring simultaneously on the computer screen \textit{(left)}, whereas individuals who are blind or have visual impairments rely on auditory cues \textit{(right)}. BLV students simultaneously listen to the teacher's voice and the screen reader vocalizing the materials. Additionally, they utilize external keyboards and keyboard shortcuts to operate the computer effectively.}}
  \Description{The image compares the differences in the scenarios of sighted individuals and BLV students using computers for learning, showcasing their distinct experiences. Individuals with sight can visually perceive multiple tasks occurring simultaneously on the computer screen (left), whereas individuals who are blind or have visual impairments rely on auditory cues (right). BLV students simultaneously listen to the teacher's voice and the screen reader vocalizing the materials. Additionally, they utilize external keyboards and keyboard shortcuts to operate the computer effectively.}
  \label{fig:teaser}
\end{teaserfigure}

\maketitle

\section{Introduction}

The significant educational inequality between people with Blind and Low Vision (BLV) and their sighted counterparts is a worldwide issue. In developing and underdeveloped regions, such as China ~\cite{Zhu2018China, Han2016Numbers} and India ~\cite{kumar2017EDUCATION, india2021Teachers}, the phenomenon of Blind Colleges,' which we define as colleges that exclusively admit BLV students for higher education, represents a unique educational model. These institutions are distinct from both `Integrated Colleges,' where BLV students study alongside sighted peers, and special education schools, which typically serve BLV students at the elementary and secondary levels \cite{Inclusive2017, morina2017inclusive, metatla2018Inclusive}. These `Blind Colleges' are shaped by historical context, transitions toward inclusive education, imperfect policies, and low capital investment, reflecting the unique social context of these areas \cite{forlin2013changing, miyauchi2022Keeping}.

Within `Integrated College,' most colleges are employing educational, accessible technology interventions to build an inclusive education context \cite{metatla2018Inclusive, hewett2020Balancing, croft2020Experiencesa, croft2021Everyone, lintangsari2020Inclusivea}. Accessible technologies such as screen readers, screen magnification systems, and refreshable Braille displays have created positive opportunities for BLV students to learn alongside their sighted peers \cite{Ariel2001computer}, including meeting diverse educational needs, improving BLV students' academic performance and professional development, promoting social integration and awareness of community diversity, and facilitating communication between BLV and sighted students in inclusive settings \cite{metatla2018Inclusive, meskhi2019Elearning, rea2002Outcomes, mag2017Benefits, Metatle2017}. By contrast, Blind Colleges often face barriers to fully leveraging these technologies, stemming from limited resources, specialized curricula, and a lack of integration with broader educational frameworks \cite{sutton2017beyond}. Such `Blind Colleges' schools indeed provide BLV students with greater opportunities to transition from elementary education to higher education \cite{chavan2023Need, prasertpong2023Factors, john2021Low, india2021Teachers}, but often within a narrow range of fields to support the sustainable personal development of individuals with disabilities \cite{alex2017quality}.

However, prior studies have primarily focused on integrated education systems in developed regions, emphasizing how accessible technologies bridge gaps between BLV and sighted students ~\cite{seale2021Dreaming, meskhi2019Elearning, croft2020Experiencesa, shinohara2022Usability}. Thus, these findings do not directly translate to `Blind Colleges,' where the socio-technical context and educational practices are vastly different ~\cite{chen2021Wasa, ng2019Shifting}, with fewer resources such as educational infrastructure, occupational opportunities, and accessible literacy, often constrained by societal factors ~\cite{Pal2016Accessibility, board2021educational}. For example, while `Integrated Colleges' emphasize inclusivity and equal opportunities, `Blind Colleges' often operate in isolation, with limited interaction between BLV and sighted individuals, which can reinforce disparities in BLV students’ exposure to diverse professional life and learning with accessible technologies. Accessible technologies are the most crucial tools for the BLV community to connect with the world. We believe it is critical to understand the current state of challenges and the use (or lack thereof) of accessible technologies in `Blind Colleges,' as well as BLV students' perceptions of them. Thus, designers, policymakers, and researchers can identify potential problems that impede successful implementation and, over time, evaluate future technologies in authentic educational contexts. Therefore, we aim to answer the following research questions (RQs) in this paper:


\begin{itemize}
    \item RQ1: What are the roles of accessible technologies supporting BLV students' learning within the `Blind Colleges' context?
    \item RQ2: How do BLV students use and perceive accessible technologies to study in `Blind Colleges'?
    \item RQ3: How do accessible technologies affect `Blind Colleges' higher education in developing and underdeveloped regions?
\end{itemize}


To address this gap, we selected China as our research context (see reasons in section \ref{Selection}) to comprehensively understand the practices, perceptions, and challenges of BLV students learning through accessible technologies in `Blind Colleges'. Due to the limited literature on the background of current Chinese BLV education, we first conducted a formative study to understand the difficulties and non-inclusiveness of the BLV education pathway in the field. Inspired by this preliminary understanding, we then conducted a two-part qualitative study (interviews and observations) in two Chinese `Blind Colleges'. Our findings highlight the importance of integrating accessible technologies into the learning environments of `Blind Colleges' while also revealing multiple practical challenges stemming from inaccessible practices. These practical challenges are similar to those faced by BLV students in `Integrated Colleges,' but their solutions are more difficult to achieve in a non-inclusive environment. Additionally, we uncovered a dilemma regarding the vision for future inclusive learning in `Integrated College', influenced by imperfect accessible technologies and the impact of long-term non-inclusive learning experiences. In our discussion, we summarized how current accessible technologies impact students' expectations toward higher inclusive education and discussed potential transitions from both technological and educational perspectives. Based on all practical challenges and future visions, we also provided several design implications for future educational accessible technologies. In summary, our paper contributes to the existing literature on inclusive education, education in developing regions, and accessible technology design by providing a comprehensive understanding of the relationship between Chinese `Blind Colleges' and accessible technologies.

\section{Background and Related Work}
\subsection{Equality, Equity, and Inclusive Education for BLV Students through Accessible Technologies}
Education equality and equity are both crucial terms in the educational domain \cite{Promoting, rachid2022quality}. These two similar terms hold different meanings; the former implies that every student receives the same resources and opportunities, while the latter means that every student receives the opportunities and resources they need based on their specific situation \cite{Equity, tprestianni2023Equity}. Inclusive education state is a higher level definition to explain this affirmation that includes all students and welcomes and supports them to learn, whoever they are and whatever their abilities or requirements \cite{Inclusive2017}. Recently, the United Nations Educational, Scientific and Cultural Organization (UNESCO) again proposed guidelines to advocate and ensure equity and inclusion in education \cite{Guide2024}, as a respected and protected right ~\cite{Right2014}. 

Thus, educational practitioners and researchers in various countries integrate students with BLV into schools from elementary \cite{Stefik2019coding, bell2019access, abramo2013Ethnographic, Haegele2016sports, Metatle2017} to higher education \cite{Lininger2008ComputersCS, hewett2020Balancing, croft2020Experiencesa, croft2021Everyone, lintangsari2020Inclusivea}, which typically involves BLV learners in classes with sighted peers. Despite the complex needs and challenges of BLV students \cite{arter1999Children, lintangsari2020Inclusivea}, the benefits of studying alongside sighted peers are overwhelming. These benefits include better grades, encouraging holistic development, promoting social integration, and increased awareness of community diversity \cite{rea2002Outcomes, mag2017Benefits, Metatle2017}. Accessible technologies are essential in this context for meeting educational needs, providing specialized resources, and facilitating communication between BLV and sighted students in inclusive settings \cite{metatla2018Inclusive, meskhi2019Elearning}.

However, many undeveloped and developing countries still have numerous `Blind Colleges,' which exclusively enroll BLV students \cite{chavan2023Need, india2021Teachers, prasertpong2023Factors, john2021Low}. These institutions are shaped by factors such as historical context, transition to inclusive education, imperfect policies, and low capital investment, which are unique to the social context of developing areas \cite{forlin2013changing, miyauchi2022Keeping}. Regarding higher education, it is also crucial for the personal development and sustainable social integration of the BLV community in developing areas \cite{prasertpong2023Factors, alex2017quality, zongozzi2022accessible}. Thus, in the effort to promote higher inclusive education, prior research also highlights the important roles of accessible technologies \cite{hewett2020Balancing, morina2017inclusive}. Yet, compared to making less inclusive areas more inclusive in well-developed regions, BLV students in developing areas face challenges of limited accessibility infrastructure \cite{Pal2016Accessibility} and lower technical literacy \cite{board2021educational}, constrain the effective use of accessible technologies. It is also widely recognized that social-technical contexts must be considered when developing and implementing educational technologies \cite{chen2021Wasa, ng2019Shifting}, indicating that experiences and challenges from `Integrated Colleges' cannot be directly replicated in `Blind College'.

All in all, the current body of research on the influence of accessible technologies on equality, equity, and inclusive higher BLV education primarily focuses on integrated universities in well-developed areas \cite{hewett2020Balancing, 2016Report, collins2005Higher, croft2020Experiencesa, shinohara2022Usability}. While this focus undoubtedly enriches the literature and provides valuable insights, it may overlook the perspectives on accessible technology and higher non-inclusive education demonstrated by BLV communities in developing societies. Our paper seeks to address this gap by revealing Chinese BLV students' insights and vision towards accessible technologies in `Blind College.'

\subsection{BLV Students Learning Through Accessible Technologies}
It is worth noting that BLV people mainly rely on their two senses, touch and hearing to access information, corresponding to touching Braille with their hands and listening to announced digital content \cite{singh2012blind}. With the rapid assistive technology, such as screen magnification systems, refreshable Braille displays, and screen readers, computers are not out of reach for the BLV people in the digital age \cite{Ariel2001computer, burgstahler2002working}. 
As a tried-and-tested solution for BLV students to interact with the world, Braille is a crucial lesson taught for each BLV individual at an early age and as an essential factor in the following academic and life achievements \cite{Judith2011BrailleChan, Silverman2018BrailleAudlts, Sheffield2022Readers, Zebehazy2014Perspectives}. Most special education school for BLV students used collage, pressed wet paper, thermoforming, and embossing to produce tactile materials with Braille as their various teaching tools \cite{Rowell2003Production, Phutane2022experience}. However, tactile material faces constant challenges with its preparation being too heavy, complex and time-consuming \cite{national2009Braille, Zebehazy2014Perspectives}, leading blind students to use the refreshable tactile display as a replacement for traditional paper-based Braille resources \cite{brewster2008display, Baker2019TVIS}. Refreshable Braille displays are a tangible interface based on densely arranged Braille dots being efficiently driven up or down to visualize images by micro-actuators, which empowers users to interact with computer interface via touch-sensitive Braille dots \cite{Chen2023Dispaly, Yang2021RTP}. While this may seem more convenient for BLV people to access information and even read images, its high cost and lack of manual customization options are notable considerations, which prevent it from completely replacing paper-based Braille.

The screen reader (i.e., VoiceOver\footnote{https://www.apple.com/accessibility/vision/} and JAWS\footnote{https://support.freedomscientific.com/downloads/jaws/JAWSWhatsNew}) is the primary assistive tool for them to access computers, which is an application that attempts to describe to the BLV user in speech what the graphical user interface is displaying \cite{singh2012blind}. The computer cursor is the key for BLV to control the screen reader and computers. They use keyboard shortcuts (such as tab and shift) to navigate the position of the focus in different areas, and the screen reader reads out the text on these, by word, by line, or by full screen \cite{Sarah2001ScreenR}. 
Apart from screen readers, screen magnification systems can magnify text and images on the computer, mainly allowing low vision users to magnify the screen area they need to see clearly through the cursor \cite{Ariel2001computer}.


Therefore, both academic and industry have made a wide range of accessibility software that can be used on the computer, such as video description \cite{Liu2021video, packer2015overview, Campos2023audio, Wang2021Automatic, Liu2021video}, image description \cite{Thapa2017Image, Lee2022ImageExplorer, Sreedhar2022AIDE, Shrestha2022Model} game playing \cite{Chavez2021games} and designing \cite{Stadler2018Gamedesign}, and efficiently coding \cite{Potluri2018Codetalk, Mountapmbeme2022Codingreview, Falase2019TactileCode, thieme2017Enabling}. When integrating those accessible technologies into educational contexts, for instance, Potluri et al. presented the CodeTalk plugin in Visual Studio\footnote{https://visualstudio.microsoft.com/} could be used in the Computer Science class for BLV students by audio cue-earcons to address the challenge of code navigation, comprehension, editing and debugging \cite{Potluri2018Codetalk, Mountapmbeme2022Codingreview}, however, whether these analogous applications are effectively leveraged to enhance the higher learning experiences of BLV students is still a concern. 
Moreover, previous studies have predominantly focused on the use of accessible technologies in integrated colleges, highlighting their benefits for fostering inclusivity. However, little attention has been given to the unique needs and challenges of Blind Colleges, where non-inclusive practices and resource constraints significantly impact the use of these tools \cite{fichten2009accessibility,  seale2021Dreaming, meskhi2019Elearning, croft2020Experiencesa, shinohara2022Usability}. Our study seeks to bridge this gap by examining how BLV students in `Blind Colleges' use accessible technologies, identifying the challenges they face, and proposing strategies to enhance their learning experiences in non-inclusive settings.

\section{Selection of Research Context}
\label{Selection}
As discussed above, a notable research gap exists in understanding how BLV students use accessible technologies in
‘Blind Colleges.’ We select China as our research context, on the one hand, China has the largest BLV community, a substantial count of approximately 17 million individuals \cite{Disabled2021number}, with the majority of students who access higher education still in `Blind Colleges' \cite{Han2016Numbers}. On the other hand, China posed a significant improvement in the practices and literacy of accessible technologies usage and facilities, which proves accessible technologies have already helped BLV people integrate into various aspects of society \cite{Liu2019shopping, Lei2022Kuaidigui, Rong2022Livestream, li2021choose}. Also, the authors conducting study and data analysis were born and raised in China. They bring rich experiences in the Chinese higher education model and are deeply familiar with accessible technologies used in the Chinese context. Combining those reasons, we select the Chinese `Blind Colleges' context as a representative social-technical context for developing countries in the following studies.

\section{Formative Study: Understand the Higher Education of BLV Students in China}
Due to the lack of relevant and lasting literature on the introduction of BLV students' higher education in China. We first conducted a formative study by visiting two Chinese special education schools. One school provides elementary education for students from 6 to 18 years old, combining primary, middle and high school, while the other school only offers higher education as a `Blind College'. Both school are exclusively enroll BLV students. We conducted multiple in-field observational studies across real classroom settings and interviewed school deans to understand their students' development plans, course syllabus, and career paths as a formative study. The formative study had three primary objectives: 1) To map the educational pathways available to BLV students in China. 2) To understand the enrollment and graduation criteria for BLV students at "Blind Colleges." 3) To explore the role of accessible technologies in teaching and learning across different educational levels.

The two special education schools we visited are all located in a northern city of China. We reached out research context through the long-term connections of the research team. On behalf of the research team, three authors participated in each in-field study. These visits involved guided tours and interviews with the school's teaching deans, followed by Q\&A sessions. The researchers then joined BLV students in the classroom, sitting at the back to observe the entire course procedure. Data collection involved taking photos, videos, and audio recordings, as well as writing notes. Afterward, two researchers transcribed the video and audio with the consent of the interviewees and analyzed that data through collaborative thematic analysis \cite{muller2014Curiosity}. Our visit was approved by the Institutional Review Board (IRB) of the research team's and both the observed institutions. Our findings specifically in three part, we found:

\begin{itemize}
    \item \textit{\textbf{1) Most BLV students choose to go special education school in China.}} In China, special education school for elementary education are typically residential, meaning students live in school dormitories during the term and have few opportunities to go out except on weekends. BLV students receive the same educational content and study the same subjects of study, including language, math, English, physics, computer, etc., as sighted students learned before entering college. After they finished their middle school studies, most blind students choose professional studies or directly enter the workforce. Only a few students continue in special education school with the aim of pursuing higher education.
    \item \textbf{\textit{2) The enrollment rates of higher `Blind Colleges' are extremely competitive}} Each year, less than about 10\% of BLV individuals, can access higher education, either as `Blind Colleges' or as part of `Integrated Colleges'. BLV students only need to take the \textit{school-organized exam}{\footnote{Separate examination and enrollment for higher professional education organized by school
    ("\textit{dan kao dan zhao}", \begin{CJK}{UTF8}{gbsn} 
     单考单招
    \end{CJK})} to compete with other blind students for `Blind College'. If they want to attend an `Integrated Colleges', they need to take the \textit{gaokao}
     {\footnote{The national undergraduate admission exam of China
    ("\textit{Gao Kao}", \begin{CJK}{UTF8}{gbsn} 
     高考
    \end{CJK}) }
    and compete with sighted people for admission. Additionally, there are only two majors available for undergraduate BLV students in `Blind Colleges': Acupuncture and Tuina Science (ATS), and Music Performance (with different instrument tracks).}

    \item \textbf{\textit{3) BLV students begin learning accessible technologies early in `Blind School.'}} BLV students start with computer classes when they first enroll in school as a children. There are teachers who teach them how to use keyboard and learn to write braille. Over time, they gradually master the use of accessible technologies such as screen readers and Braille displays during their study and life. As a result, when they enter college, they are generally equipped with a foundation in learning and understanding how to operate accessible software.}
\end{itemize}

In conclusion, our formative study objectively illustrates the background of learning with accessible technology in Chinese BLV education. These findings provide us with a preliminary understanding of the difficulties Chinese BLV students face when enrolling in and studying in `Blind Colleges'. All this information motivates us to conduct further studies. We believe that understanding their challenges, perspectives, and experiences with using accessible technology for learning could provide meaningful inspiration for future BLV education and foster inclusion in the current non-inclusive educational environments.

\section{Methods}
Based on our formative study and literature review, we conducted two-part human studies with qualitative analysis to answer our research questions, including an observation study (N=9) and an interview study (N=20). Our research methods are similar to methods used in previous research about understanding accessible technologies of BLV communities ~\cite{Rong2022Livestream, Li2021Cooking, Li2022Makeup, huang2023Exhibition, Lei2022Kuaidigui, Liu2019shopping}. Our studies have been approved by the last author's Institutional Review Board (IRB). All participants provided consent before the researcher did any data collection.
\begin{figure}[H]
    \centering\includegraphics[width=\linewidth]{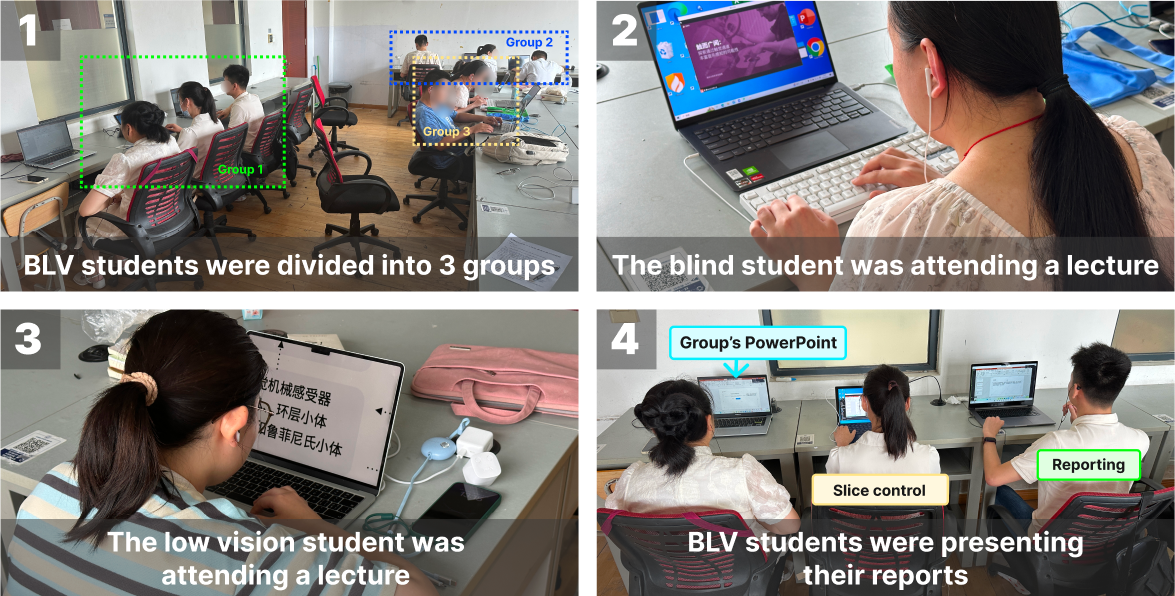}
    \caption{These figures illustrate the overview of an in-field mock-up workshop. (1) BLV students were divided into three groups. (2) One BLV student attended a lecture (3) One low-vision student attended a lecture. (4) one group of BLV students were presenting their reports
}
    \label{fig:Observation overview}
    \Description{This figure contains four images. The first image depicts BLV students sitting in a classroom, divided into three groups, with three students in each group. In the second image, a blind student is wearing headphones and using an external keyboard to operate a computer. An online conferencing software is open on the computer screen, displaying the teacher's lecture notes. The third image shows a low vision female student reading the teacher's lecture notes on a computer, with the notes enlarged for better visibility. The fourth image depicts three BLV students giving a group presentation, with one student speaking while another operates the PowerPoint slides to complement their presentation content.}
\end{figure}

\subsection{Observation Study Design}

To comprehensively understand our research questions, we conducted an observational study to objectively explore students' use of accessible technologies during their in-class learning \ref{fig:Observation overview}.

\subsubsection{Observation Participants Recruitment}
The student participants of the observation study were recruited from a Chinese `Blind Colleges'. Participants' `Blind Colleges' have, and only have, two majors for BLV students to pursue, which is Acupuncture and Tuina Science (ATS), and Music Performance. We first contacted this person through a faculty member at `Blind Colleges' in partnership with the research lab and asked him to use a snowball sampling method within his campus network to recommend BLV students who met the criteria for the experiment and to request that these students invite their classmates \cite{goodman1961snowball}. The criteria are met by `Blind Colleges' students who are familiar with accessible technologies. The detailed participants' demographic information regarding the student's age (M = 24.1; SD = 1.70), gender (4 Male; 5 Female), degree of visual impairment (8 totally blind; 1 low vision), field of study (8 Acupuncture and Tuina Science (ATS); 1 Music) and educational level (5 Undergraduate; 4 Graduate), is provided in table \ref{Participants’ demographic information.}. Each participant signed a consent form for their involvement and received compensation of 300CNY for their time.

\begin{table*}[h]
\caption{Workshop Participants’ demographic information.}
    \centering
\begin{booktabs}{
  cell{2}{1} = {c=6}{},
  cell{6}{1} = {c=6}{},
  cell{10}{1} = {c=6}{},
  width=\linewidth,
  colspec={XXXccc},
  cells={c},
}
\toprule
No.    & Age  & Gender & Visually impaired degree               & Major & Education background \\
\midrule
Group1 &      &        &                                        &       &                      \\
\midrule
S1     & $23$ & Female & Congenital blindness                   & ATS   & Graduate             \\
S2     & $25$ & Female & Congenital blindness                   & ATS   & Graduate             \\
S3     & $23$ & Male   & Acquired blindness, since 7 years old  & ATS   & Undergraduate        \\
\midrule
Group2 &      &        &                                        &       &                      \\
\midrule
S4     & $21$ & Male   & Congenital blindness                   & Music & Undergraduate        \\
S5     & $26$ & Male   & Congenital blindness                   & ATS   & Undergraduate        \\
S6     & $26$ & Female & Acquired blindness, since 9 years old  & ATS   & Undergraduate        \\
\midrule
Group3 &      &        &                                        &       &                      \\
\midrule
S7     & $25$ & Male   & Congenital blindness                   & ATS   & Graduate             \\
S8     & $23$ & Male   & Acquired blindness, since 13 years old & ATS   & Graduate             \\
S9     & $25$ & Female & Congenital low vision                  & ATS   & Undergraduate         \\
\bottomrule
\end{booktabs}
    \label{Participants’ demographic information.}
\end{table*}

\subsubsection{Observation Study Procedure}
\begin{figure}[h]
    \centering
    \includegraphics[width=0.5\textwidth]{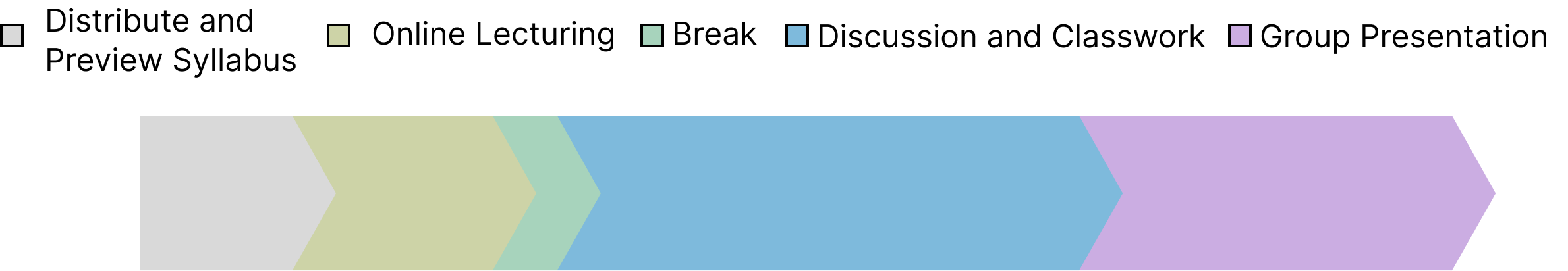}
    \caption{\textbf{This flowchart illustrates the procedure of a mock-up course workshop with the approximated time. }}
    \Description{The picture shows the flowchart for this Workshop procedure arrangement. The flowchart consists of a rectangle and some markup. The rectangle shows the duration of each session of the workshop, and the name of each session is labeled below the duration. From left to right they are: Preparation and Preview Coursework (15 minutes), Online Class (20 minutes), Break (5 minutes), Discussion and Group Work (60 minutes), and Presentation (40 minutes).}
    \label{fig:Flowchart}
\end{figure}

We collaborated with an instructor to design a mock-up course workshop aimed at providing a comprehensive in-class learning process, including a preview syllabus, lectures, group work, and presentations (see Figure \ref{fig:Flowchart}). Our mock-up course effectively simulates a real learning scenario, as BLV students often participate in online guest lectures and are familiar with online classrooms due to their experiences during the COVID-19 pandemic. We first distributed the syllabus document and slides, introducing the workshop’s agenda and the lesson context. Participants were also given instructions on downloading virtual meeting software. Following that, participants were connected to a virtual meeting room, where the instructor delivered a 30-minute online lecture on novel knowledge to BLV students, interdisciplinary design for accessibility, and HCI. In the third session, students were grouped into three and assigned tasks to collaborate on, and a presentation was created demonstrating their learning practice in summarization, research, and collaboration (refer to Appendix \ref{Classwork} for detailed task instruction). Each group would deliver their finding, with presentations lasting approximately 10-20 minutes each. The process of the workshop could be visualized in Figure \ref{fig:Flowchart}, including each of the four steps and the approximate time for each section. Each section is color-coded with different colors.

\subsection{Interview Study Design}
To comprehensively understand our research questions, we conducted multiple one-on-one interview studies to gain deep, diverse, and complex insights into students' perspectives on the underlying reasons for their social behaviors and the challenges they face when using accessible technologies in both in-class and out-of-class learning \cite{carpendale2017Analyzing}.

\subsubsection{Interview Participants Recruitment}
Our participants were BLV students from two `Blind Colleges' in China, recruited through social media and a long-term partnership between the `Blind College' and our research lab, using a snowball sampling method recommended by previous BLV participants ~\cite{goodman1961snowball}. All were native Mandarin speakers with varying degrees of visual impairment. Table \ref{table2} illustrates their detailed demographic information, including age ranges from 19 to 27 years old (M=22.35; SD=2.30), gender distribution (8 male; 12 female), majors (15 in Acupuncture and Tuina Science (ATS); 5 in Music), and educational levels (17 undergraduates; 3 graduates). Twelve of the participants were interviewed in person in their school classrooms, while the remaining eight participated in interviews conducted over Zoom with computer screensharing.

\begin{table*}[htbp]
\caption{The demographic information of interview participants.}
\label{table2}
\Description{This table shows the demographic information of 20 interview participants in ages, gender, visually impaired degree, major education background}
\renewcommand{\arraystretch}{1.0}
\begin{booktabs}{
  colspec={X[c]X[c]X[c]ccc},
  width=\linewidth,
}
\toprule
No. & Age & Gender & Visually impaired degree                      & Major & Education background \\
\midrule
P1                                 & 24  & Male   & acquired blindness, since 17 years old        & ATS   & graduate             \\
P2                                 & 22  & Male   & acquired blindness, since 2 years old         & ATS   & undergraduate        \\
P3                                 & 23  & Male   & acquired blindness, since 8 years old         & ATS   & undergraduate        \\
P4                                 & 27  & Male   & acquired blindness, since 7 years old         & ATS   & undergraduate        \\
P5                                 & 19  & Male   & acquired blindness, since 17 years old        & Music & undergraduate        \\
P6                                 & 19  & Male   & congenital blindness                          & Music & undergraduate        \\
P7                                 & 21  & Female & congenital low vision                         & Music & undergraduate        \\
P8                                 & 23  & Male   & congenital blindness                          & ATS   & undergraduate        \\
P9                                 & 23  & Female & congenital blindness                          & ATS   & undergraduate        \\
P10                                & 25  & Female & congenital blindness                          & ATS   & graduate             \\
P11                                & 26  & Female & acquired light perception, since 12 years old & ATS   & graduate             \\
P12                                & 22  & Female & congenital low vision                         & ATS   & undergraduate        \\
P13                                & 22  & Female & acquired low vision, since 6 years old        & ATS   & undergraduate        \\
P14                                & 21  & Female & congenital low vision                         & ATS   & undergraduate        \\
P15                                & 22  & Male   & acquired blindness                            & ATS   & undergraduate        \\
P16                                & 21  & Female & acquired blindness                            & Music & undergraduate        \\
P17                                & 26  & Female & acquired blindness, since 11 years old        & ATS   & undergraduate        \\
P18                                & 20  & Female & congenital blindness                          & Music & undergraduate        \\
P19                                & 20  & Female & acquired blindness, since 3 years old         & ATS   & undergraduate        \\
P20                                & 21  & Female & congenital blindness                          & ATS   & undergraduate    \\  
\bottomrule
\end{booktabs}

\end{table*}

\subsubsection{Interview Design}
Our interview protocol followed a three-part semi-structured format, and we asked follow-up questions when participants' answers lacked detail or were intriguing. Before beginning the interviews, we asked participants for permission to audio and video record their responses. We started each interview by collecting demographic information and asking participants to describe their daily learning schedule, detailing their learning needs and relevant choice of accessible technologies reasons. Next, we asked each participant to explain their practices in the self-directed learning process. We also asked them to show us their frequently used learning software or functions on their accessible technologies, and we observed or recorded (if they approved) any challenges they faced while using these tools. When participants attempted to fix these issues, we also asked them to think aloud about their perceptions of the interface design, their explanations of the practices, and their expectations for possible solutions. Finally, we asked participants about their feelings and experiences of using accessible technology and how they perceive the connection between their current learning opportunities and vision of future inclusive education. Each one-on-one interview lasted approximately 25 to 35 minutes in Mandarin, and participants were compensated equally for their time. The detailed interview protocol can be found in Appendix \ref{Interview}.

\subsection{Data Collection and Analysis}
Throughout the workshop and interviews, with the participants’ consent, we video-recorded the entire study procedure, including on-course listening, collaborative study, presentation, and operations on assistive technology for study. We also audio-recorded all focus group discussions and interviews.

Our data analysis was guided by the thematic analysis method \cite{braun2012thematic}. We initially analyzed the 2 hours of observation video and 11 hours of interview transcripts using the open-coding method to generate initial codes \cite{khan2014qualitative}. For interview transcription, the two authors, native Chinese speakers, and fluent English speakers, first used a voice transcription platform to transcribe all recordings into English and manually resolved grammatical errors. For the video transcription, we particularly emphasized identifying actions by participants that were related to our research goals. Two researchers first performed open-coding of the transcripts and videos using MaxQDA\footnote{Qualitative data analysis software, https://www.maxqda.com/} individually. They then exchanged codes and discussed their perceptions of various words or clips to address conflicts of personal bias. After achieving consensus on their codes, the researchers conducted an affinity diagram method to group the codes into clusters and summarize emerging common themes with internal connections.


\section{Findings}
In this section, we provide a comprehensive analysis of the practice and challenges of BLV students using accessible technologies to studies in the `Blind Colleges' context, which reflects the unsatisfactory and non-inclusive educational experiences and provides insights into designing better accessible technologies as the basis of future inclusive education. We first presented an overview of how BLV uses various accessible technologies to meet their in-class and off-class learning needs in Section 5.1. After that, in Section 5.2, we list four primary challenges of accessible technologies that hinder BLV students' learning. Section 5.3 further explores how our participants perceive current accessible technologies and how they impact `Blind College.' Finally, section 5.4 reported how BLV students expect future accessible technologies could foster their higher inclusive education and provide concern about this process.

\subsection{Overview of Accessible Technologies Learning Practice}

In general, we found that BLV students primarily rely on digital and tactile accessible technologies to enrich their learning experiences and gain access to educational resources. In this subsection, we only provided an overview of why and when to use various accessible technologies for what learning purpose to avoid redundant descriptions. 
\subsubsection{Digital Accessible Technologies Practice}
\begin{figure}
    \centering
    \includegraphics[width=1\linewidth]{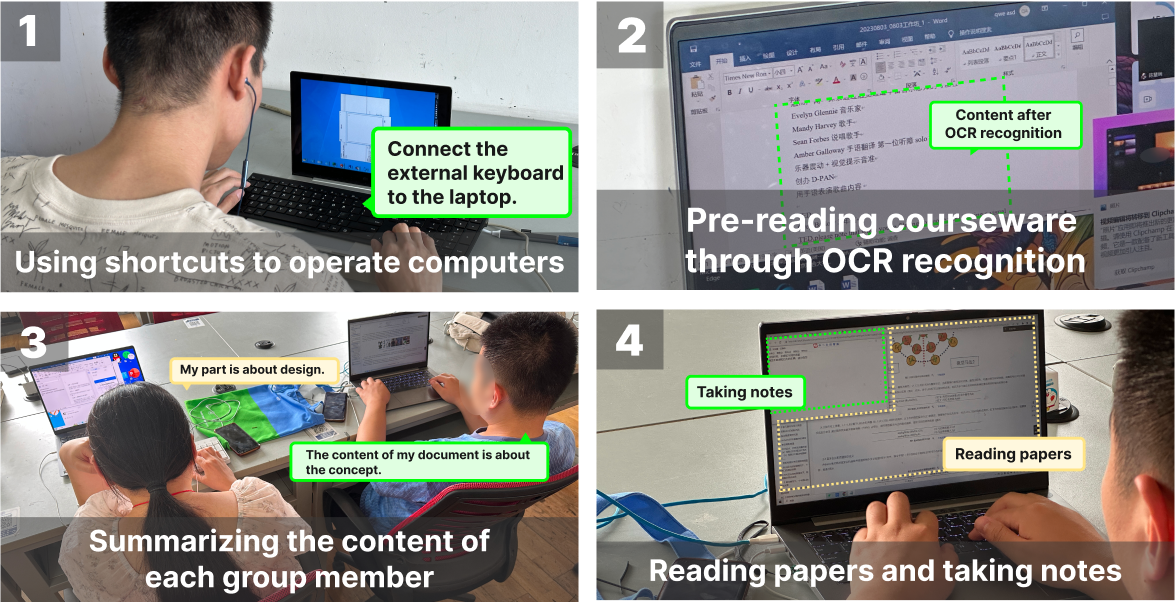}
    \caption{These figures show key features of how BLV students use digitally accessible technologies for learning, such as 1) using shortcuts to operate computers, 2)pre-reading courseware through OCR recognition, 3) collaboratively summarizing contents of reading by each group member, 4) Reading the paper and taking notes at the same interface.}
    \label{Practice}
    \Description{This figure contains four images. The first image shows a BLV student wearing headphones and using an external keyboard's shortcuts to operate the computer. The second image illustrates a visually impaired student using OCR recognition to read course materials. In the third image, a male BLV student and a female BLV student are seen discussing how to collaborate. The male student says, "The content of my document is about the concept," while the female student adds, "My part is about design." They are trying to summarize the information they have gathered. The fourth image depicts a BLV student with two computer windows open. One window displays a webpage used for reading literature, while the other window shows a notebook software used to record important content from the literature.}
\end{figure}

We found that BLV students primarily utilize computers and smartphones as digital accessible technologies. Among these, computers serve as their primary learning tool and are noted as the most efficient for educational purposes. Some participants, such as P6, augment their computers with external keyboards to facilitate faster typing and more efficient software navigation.
In contrast, smartphones are primarily used to meet daily life needs and are considered inefficient for learning. Yet some still utilize them for specific learning purposes, such as P11 and P14 who use the smartphone camera to recognize text. 





In terms of software technology, BLV students most commonly utilize basic built-in software for their studies rather than specially designed software. For instance, some of our participants use Word and Notes for note-taking, while P2 utilizes Office software to compile and organize materials (Fig.\ref{Practice}-4). Another indispensable tool that serves as a powerful aid for BLV students in using computers seamlessly is the screen reader. P4 believes that the screen reader can effectively assist in non-graphical learning, enabling him to operate the computer without obstacles. They use OCR recognition to read the content for inaccessible graphical resources, such as figures and tables (Fig.\ref{Practice}-2). BLV students also show high interest in learning knowledge beyond their majors by accessing online resources, including electronic textbooks, electronic materials for assignments, and online videos for learning various subjects. When BLV students need to collaborate with each other, they primarily divide the work into individual tasks and integrate everything at the end (Fig.\ref{Practice}-3).

\subsubsection{Tactile Accessible Technologies Practice}
We also found that BLV students still use accessible tactile technologies (mostly small-scale refreshable Braille display) alongside accessible digital technologies for some particular learning needs such as studying subjects like English that require understanding and memorization. The main reason is that they can read Braille materials prepared for the BLV community through these technologies, and tactile learning enables them to achieve better memory retention. It also helps BLV students better learn the spelling of English words because each letter is displayed in Braille individually. As P8 stated:
\begin{quote}
\textit{``...using screen reader software to read English words letter by letter is very difficult for comprehension. When trying to memorize a word, feeling the Braille dots on the display can make it easier to remember the word than simply relying on auditory memory alone...'' — P8}
\end{quote}

For BLV students who major in music, they also use Braille displays to read sheet music. This is not only because they can use Braille displays to read Braille music notation but also because they typically memorize the music they perform. Therefore, feeling the Braille sheet music is more convenient for memorizing the score, and there are Braille sheet music resources online that facilitate extended reading using Braille displays. However, our participants also mentioned braille displays are gradually falling out of use due to the rise of e-learning resources and their limited accessibility.

\subsection{Challenges Hinder Learning Through Accessible Technologies} In this subsection, we identify four primary challenges hindering BLV students' learning: unperceived graphic content (Section \ref{Unable Perceived Graphical Learning Information}), limited accessible resources (Section \ref{Shortage of Accessible Resources Leads to Cumbersome Learning Processes}, lack of software accessibility considerations (Section \ref{Non-accessibility Considerations Results in Disruption of Learning}), and low accessibility accuracy (Section \ref{Low Accessibility Accuracy Leads to Inefficient Learning}), therefore, these issues lead to inefficient, cumbersome, and disrupted learning processes.
\subsubsection{Unable Perceived Graphical Learning Information}
\label{Unable Perceived Graphical Learning Information}
Based on our interviews and observation studies, we discovered that inaccessible visual learning information still poses significant challenges for BLV students at `Blind Colleges', despite the proliferation of image accessibility technologies. As P19, a student majoring in ATS, expressed, accessible technologies meet her 90\% learning needs, except in the category of graphical aspect and visual information.

BLV students often need help understanding image-related learning content, such as textbook illustrations. Typically, this requires explanations from a sighted person or a teacher through detailed descriptions, and BLV students repeat questions and answers about what needs to be clarified. Yet, BLV students still express difficulties in fully visualizing the content. For example, P11, a student majoring in ATS, shared his challenges in learning about medical imaging: \textit{``...For the imaging is something that we can only look at, like the report of x-rays, we can not see the certain images. The direct consequence is that it creates a significant problem for our diagnosis during major learning...''}

\begin{figure}
    \centering
    \includegraphics[width=1\linewidth]{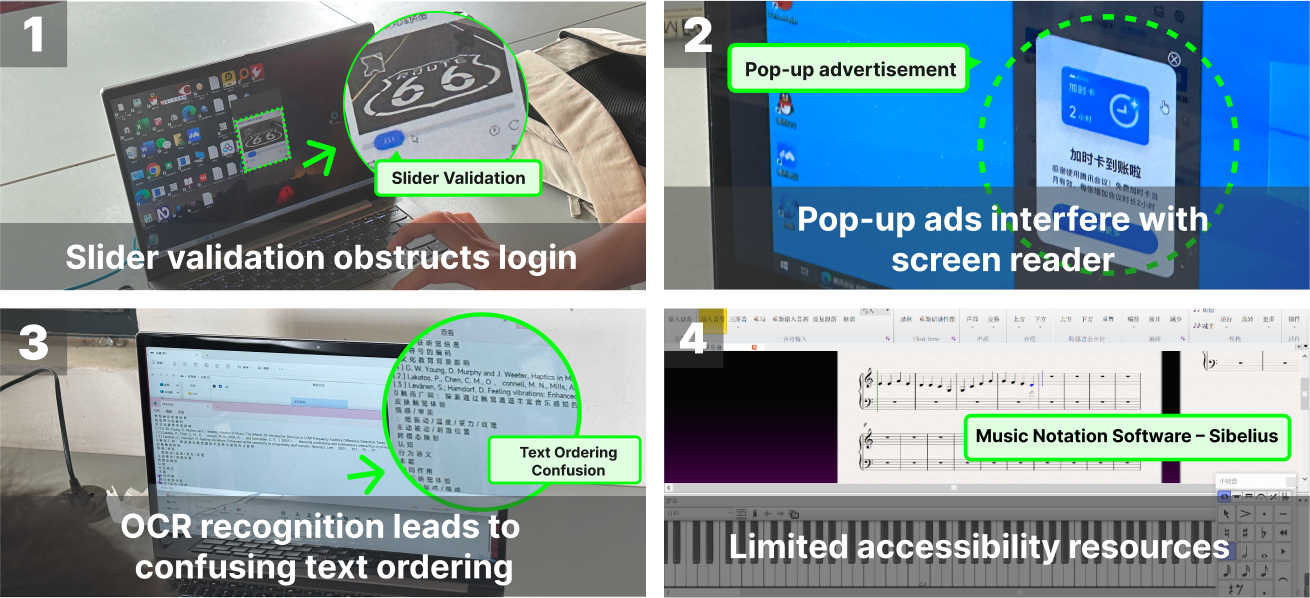}
    \caption{These figures visualize the challenges encountered by BLV students when using accessible technologies, including 1) Graphic operations such as slider validation that obstructs login; 2) Suddenly appearing pop-up ads during software usage; 3) OCR recognition errors that result in confusing text order; 4) The necessity for BLV students to use music notation software like Sibelius to create their own learning resources.
}
    \label{challenges}
    \Description{This figure contains four images. The first image illustrates BLV students unable to use slider validation to log in to software. The second image shows pop-up ads appearing during the login process, and BLV students are unable to recognize these pop-up ads using screen reading software, resulting in usability difficulties. The third image depicts BLV students using OCR to recognize documents, but encountering garbled text that prevents them from reading the file properly. The fourth image portrays limited accessibility resources, as BLV students attempt to use a music software called "Sibelius" to read musical notation, since there are no braille scores available for them to study.}
\end{figure}

Apart from the visual information in learning materials and textbooks, BLV students often encounter difficulties with graphical aspects during the learning process. For instance, when browsing web-based learning materials, they frequently face disruptions from advertising messages and P10 and P11 mentioned that they have to listen and read carefully before deciding whether to skip certain information (Fig.\ref{challenges}-2). Moreover, interactions involving sliders and pictures, such as CAPTCHAs, prove challenging. Consequently, BLV students typically seek assistance through services like \textit{`Be My Eyes \footnote{https://www.bemyeyes.com/}'} or enlist the help of sighted acquaintances. Take P1 as an example; he mentioned that BLV students find it impossible to manage such challenges independently, which significantly hampers their learning (Fig.\ref{challenges}-1). Indeed, our participants acknowledge the advancements in accessible technologies for image reading. However, they share concerns about the difficulty of learning to use these tools, as neither their peers nor teachers are familiar with them.

Moreover, our findings illustrate that BLV students also face challenges when they need to output visual information during their learning, such as collaborating with others, making presentation slides, and editing documents and fonts. Based on our observation study, we found that BLV students have to ask sighted people or low-vision classmates to help with this process. Yet, P1 presented his concern as \textit{``...If I talk to him through his language, I will probably know what's going on. But if you ask me if (what I think and what he does) is exactly the same as what he does, I don't know...''} In addition, P4 shared his difficulties in the practice of learning coding independently and collaboratively:
\begin{quote}
  \textit{ ``...I think the slightly bigger difficulty is probably drawing diagrams, and when you're communicating with other people you might use flowcharts, right? Yeah, you can't draw a flowchart, or you can draw a flowchart, but you can't read it. If you use (some applications) to draw a diagram, you can write it in code and then render it as a diagram, and people can read it, but I can not...(P4)''}
\end{quote}


\subsubsection{Non-accessibility Considerations Results in Disruption of Learning}
\label{Non-accessibility Considerations Results in Disruption of Learning}
We also discovered that many common and professional software lack accessibility considerations, posing significant obstacles for BLV students. In 2023, when interviews were conducted, specialized software such as statistical tools and collaborative writing platforms in China still do not support Chinese screen readers and other accessibility tools. For example, P5 highlighted problems with music sheet reading applications that lack interaction, making it difficult to follow along while playing. In addition to these professional tools, there are common applications also remain inaccessible, as P9 explained:

\begin{quote}
\textit{``...For example, my cloud disk... Sometimes it's not that the screen reader can't read the information, but the software development didn't consider screen reader users, leading to mismatches...''(P9)}
\end{quote}

BLV students acknowledge that accessible technologies alone cannot fully address their challenges, given their heavy reliance on common software. Issues like unreadable content, unresponsive shortcut keys, and inaccurate image text recognition persist, affecting their access to learning resources. Specifically, when seeking self-directed learning materials online, they encounter videos not tailored for their needs. For instance, when P4 self-learning coding online, he said, \textit{``...For instance, the instructor might use image illustrations or display code directly on the screen, but it's unreadable... Then, I'm trying to use VS Code editor, and the instructor says 'click on the green button, click on the red button,' and I'm lost...''} In situations like this, BLV students feel that their learning needs remain unmet when creators fail to prioritize accessibility. This often happens because creators are unaware that a group of BLV students is using these materials for learning.

Apart from that, some software lacks accessibility considerations despite having accessibility functions. P6 and P19 face challenges when taking notes while reading literature. They find the annotation functions used by sighted individuals inconvenient, as they require selecting the original text before performing actions. This, coupled with complex structural logic, makes locating the typing area difficult. P6 converts literature into Word documents and divides articles into sections to record keywords and thoughts, but this often results in mixing up the original text with notes. On the other hand, P19 creates separate documents for note-taking, but this prevents her from associating her notes with the original content.

\subsubsection{Shortage of Accessible Resources Leads to Cumbersome Learning Processes}
\label{Shortage of Accessible Resources Leads to Cumbersome Learning Processes}
Throughout our interview study, participants agreed that the shortage of accessible resources in `Blind College' creates challenges in their learning processes and complicates them. Finding such resources, both online and offline, is difficult, and converting personalized materials into accessible formats is tedious. Beyond textbooks, additional resources for professional or self-directed learning are often inaccessible from their school. Students currently only rely on recommendations from peers and instructors to sidestep the burdensome production process. Otherwise, their manual production of accessible resources would require a cumbersome process, as P3 fears,

\begin{quote}
  \textit{``...Translating all market textbooks into Braille or electronic formats is a massive task. Even if you do, important visual information, like diagrams for physics, chemistry, and math experiments, might get left out....(P3)''}
\end{quote}

To address this issue, P1 described the process and strategies he uses when he wants to read a book: \textit{``...A Braille book is essentially translated from Chinese characters into electronic Braille, and then... in the Library for the Blind, there is a post specifically for manual proofreading, so they need to correct it manually before it is engraved and printed into a book..."} However, this process is equally cumbersome, requiring BLV students to prepare resources themselves and then rely on sighted individuals for proofreading, which doesn't meet their immediate learning needs because they generally don't have sighted peers to assist in.

Moreover, BLV students are also struggling to create their personalized accessible resources, especially in fields like music (Fig.\ref{challenges}-4). P5 explained, \textit{``...Different publishers may have versions of the same piece, making comparisons difficult...''} Consequently, some students, like P18, resort to listening to audio online and transcribing it into piano sheet music: \textit{``...I usually find the audio on the internet, play it at 0.5x speed, and listen to it bit by bit. I can't guarantee that it's completely accurate, but I can at least get a rough idea of what it sounds like..."} Personalized learning is highly valuable for every learning process; however, such a process prolongs BLV students' study time.

\subsubsection{Low Accessibility Accuracy Leads to Inefficient Learning}
\label{Low Accessibility Accuracy Leads to Inefficient Learning}
Based on our interview and observation studies, we found that BLV students' efficient learning would be interrupted by low accessibility accuracy rather than the proficiency factor in the operation of accessible technologies. The low accessibility issues mainly stem from text recognition errors, homophone pronunciation errors, punctuation errors, and art fonts that can not be recognized, etc. For instance, BLV students frequently need to read PDFs that contain scanned images of texts for their learning materials. They rely on software such as OCR, to transcribe the text from the images or PDFs into Word, but the accuracy of these methods is low, and they often disrupt the original formatting (Fig.\ref{challenges}-3). For instance, P6 said:
\begin{quote}
\textit{``...For example, if we go to the piano class or the erhu or whatever. We can't see electronic sheet music unless it is specialized with the kind of sheet music software to type out, now a lot of online sheet music is that kind of picture, then this time, we may use that OCR recognition, but also not quite be able to recognize it, on the OCR recognition is basically no use of a state, and this time it is more difficult...(P6)''}
\end{quote}
Apart from that, low accessibility accuracy can cause confusion and difficulties in understanding the content of original learning resources. In the process of learning English, BLV students found the pronunciation mechanics read out by screen readers difficult to understand. P9 shared her unsatisfactory learning experiences due to low accessibility accuracy in learning English \textit{``...the original Chinese content is already (hard to read)... no matter what more things, knowledge symbols what can not be recognized is already very annoying. Well, English is not to mention. Aigoo, it's a series of (error) I really do not know how to (do)...''}

\subsection{Perceptions of Learning through Accessible Technologies in `Blind Colleges'}
\label{sec:findings-impacts}
In this subsection, we report how BLV students perceive their non-inclusive `Blind Colleges' learning experience, feeling uneven and unequal regarding educational and accessible resources. Beyond those negative expressions, we also found that BLV students understand their own shortcomings and believe these stem from the non-inclusive educational paths they have followed since an early age. Moreover, their unique pathway created a huge practical gap in the usage of accessible technologies.

\subsubsection{Uneven and Unequal Learning Experiences}
We found that BLV students generally had too few higher education majors to choose from and wished that more options could be made available. Additionally, there is significant unevenness within different cities of the country; `Blind Colleges' is also not an easily accessible resource for low-income cities and areas. This issue is not exclusive to `Blind Colleges' but is also true for the small number of students who can attend an inclusive school, where the choices of majors remain very limited in liberal arts rather than STEM major. As P9 said, \textit{``...We can choose ATS and Music at our college, and at other `Integrated Colleges', it's basically the same, maybe with the addition of applied psychology or special education. But that's it; there's not much more (options)...''}. P20 shared a similar view regarding the limited selection of majors, which did not match students' personalized needs, stating:

\begin{quote}
\textit{``...I think there are still too few specialties. ATS is a very good thing, after all, it is part of our traditional Chinese culture and is very good, but not everyone is suited for it... Some girls may not have enough strength and may not be suitable for this major...(P20)''}
\end{quote}

It is worth noting that, even though BLV students recognize the inequality, they understand that this doesn't mean they are completely helpless if they wish to study at an `Integrated College', which is an equitable opportunity from society. For example, P1 shared his thoughts: \textit{``...If you said that this person is also very talented, then they can totally go for (the best college), right? And we know people who are going for a PhD; there are many, as long as you have the ability, you can totally do it...''} However, the reality is that few Chinese BLV students choose this path, and our following sub-subsection illustrates the underlying reasons.


\subsubsection{Influenced by Non-Inclusive Education from an Early Age}
BLV students believe there is a crucial factor leading to their failure or avoidance of the \textit{gaokao}, choosing to attend `Blind Colleges' instead: their elementary education in special education school is easier, low-pressure, and more lenient, leaving them uncompetitive when taking the same exams as sighted students. We found that, due to BLV students having a specific higher education entrance path, which \textit{``...leads directly to a school-organized exam, so the underlying logic is that BLV students aren't stressed, we don't need to learn as much...''} as stated by P9. This situation leads to a scenario where, even though BLV students would like to pursue higher studies in an `Integrated College', it's too late to make that decision in high school. P20 also proposed her idea about why there is less stress for them: 
\begin{quote} 
\textit{``...Maybe the teacher doesn't think it's necessary, because it's really difficult to talk about geometry and stuff like that... I guess it may not be that the teacher himself feels he does not attach importance to this matter, he hasn't thought deeply about the knowledge, he may think that everyone is going to a school-organized exam which isn't that difficult...(P20)''}
\end{quote}

Continuing into higher education, P12 detailed their current stress even at the college level: \textit{``...We don't seem to have a lot of homework. Well, right now it's just basically a verbal test from the teacher and that's about it...''} They believe the course content is not overly in-depth and is primarily designed to cater to the needs of the majority of students.

\subsubsection{The Diverse Capability of Learning Accessible Technology within `Blind Colleges'}
Despite the insights we gained from Section 3 and the interview, BLV students are taught to use accessible technology during their time in elementary special education school. P10, for example, shared, \textit{``...When I was in middle and high school, we had computer classes, and then I was taught a little bit (of operation) in every class at that time. I learned it slowly...''} We found that there was still a wide gap in the level of accessible technology usage among our participants. Many students are not familiar with accessible technologies because they do not fully understand their importance before entering 'Blind Colleges' with low technical literacy. For example, P8 mentioned that some students, like herself, only began officially learning computer skills upon entering college:

\begin{quote}
\textit{``...When I was in primary school, I was not interested in computer classes and I didn’t perceive computers as particularly useful since I primarily relied on paper-based Braille materials for studying. However, when I entered college and realized that almost everyone had a laptop, I understood the importance of computers as a learning tool. So, I had to start learning from scratch on my own...(P8)''}
\end{quote}

Many participants emphasized the importance and challenges of mastering keyboard shortcuts in computer-based learning. Proficiency in using shortcuts requires consistent practice and repetition to memorize them gradually. P14 mentioned that learning keyboard keys was easy, but executing various functions using the keyboard posed difficulties. P1 elaborated on a specific illustration of his computer learning process, \textit{``...I started using computers in sixth grade... However, it was only after I entered college, when I had my own personal computer, that I truly became proficient and utilized it effectively to support my studies...(P1).''} Therefore, some students new to educational accessible technology may feel embarrassed when they study alongside their peers. P8 shared that:

\begin{quote}
\textit{``...When I first came to college, the teacher said that taking notes in class... I went over there with a Braille paper and a Braille pen, while everyone was typing and I was the only one taking notes in Braille, and I felt like all the students were looking at me like I was an idiot. I was so embarrassed, because I felt so outdated...(P8)''}
\end{quote}

\subsection{Future Vision of Accessible Technologies in Higher Inclusive Education}
In previous findings, we found that accessible technologies play a significant role in the current education system at `Blind Colleges', with the unsatisfactory state of both accessible technologies and the non-inclusive environment at `Blind Colleges'. This subsection separately reported what our participants shared about their expectations and concerns for improving these two aspects of their educational experiences.

\subsubsection{Dilemma of Future Inclusive Learning Experiences}

As discussed in Section ~\ref{sec:findings-impacts}, BLV students in China wish to have more major selection choices and opportunities to study alongside sighted peers in an inclusive campus environment. Students believe that society and educational staff should take responsibility for guiding BLV students in choosing majors and informing them about their options to help build their worldview, rather than \textit{``...opening a `Blind College' to give BLV students customized training, which in fact may limit some people's talents...''} as P1 stated. P6 shared his similar idea from the perspective of after-graduation life, \textit{``...Although most of us will choose these types of special education universities, we are ultimately geared towards the able-bodied community, and there is a certain disconnect between (now)...''} 

However, on the other hand, BLV students express concerns within the Chinese social context that teachers in higher education generally lack accessible literacy and experience in teaching BLV students. Our participants shared that their learning methods are completely different from those of sighted students, which can be difficult for teachers, who are sighted themselves, to imagine and adjust the pacing of their lectures accordingly. For example, P18 expressed her concerns:\textit{``...There's also the fact that a teacher can't just have one or two BLV students in a class and then slow down because of those one or two BLV students. Yeah, so I think there's a practicality to it...''}

Apart from that, we found that the gap between congenital conditions and reality may pose a concern for BLV students in their inclusive education. Meanwhile, our participants felt that although accessible technology could help them to some extent to join in inclusive education, it was still not sufficient to compete with professional sighted individuals in both academia and broader society. For example, P20 said, \textit{``...I used to want to go into voice acting, but I thought I might need to rethink and be more cautious... The reason is that you have to compete with those professional sighted people, which is inevitably very difficult, much more difficult than if you stay within the system of the visually impaired group...''}. P18 expressed similar ideas about his current major, music:
\begin{quote}
\textit{``...The main thing in music is the score, and listening to the score as I do is definitely much less efficient than reading the score, you have to admit that... Ultimately, when it comes to playing, you have to memorize the score. Yes, you have to spend a lot of time memorizing the score, which is another delay. While you are memorizing the score, people are getting ahead of themselves, which is also a very practical problem...''}
\end{quote}

\subsubsection{Non-Specialized Design of Accessible Technologies}
\label{Non-Specialized Vision of Accessible Technologies}
Regarding accessible technologies, our participants commonly believe they do not wish for educational, accessible technologies to be specialized for their unique learning habits but rather to remain the same as those used by sighted people. Many accessible technologies seem designed to be convenient for the BLV community. However, this actually increases their usage pressure. For example, P4, a blind student, shares his concern: \textit{``...Another aspect is that it's hard to get in line with the mainstream. For instance, I'm using VS Code in Google and can probably find a solution. However, if something goes wrong with an IDE developed specifically for blind people, or if I don't know how to use a feature, I can't search for it at all...''}. This perception is also caused by the general tendency of BLV students to avoid troubling others, instead demonstrating remarkable patience and perseverance in attempting to independently solve problems until they have no choice but to seek help from sighted individuals. For example, P20 expressed,

\begin{quote}
\textit{``...Well, it seems like other students, too, ask the teacher, but many of them tend to solve things on their own first. Maybe it's a habit of ours (BLV students). I think we don't want to bother others and try to solve the problem by ourselves...(P20)''}
\end{quote}

BLV students also express concerns that specialized accessible technologies might be abandoned due to their limited market appeal. Instead, BLV students wish for the integration of more subtle accessibility features into current screen readers. For example, P10 explained the current imprecise reading experience when trying to locate specific content: \textit{``...If you want to listen to a certain word or a certain sentence, there’s no way to jump directly to it. You can only navigate one page at a time, not sentence by sentence...''}. Regarding accessible technology for haptic-assisted learning, our participants expressed a desire to utilize large-scale refreshable Braille displays to illustrate visual-related content better and enhance reading for subjects like geometry, sheet music, and medical images because they perceive tactile senses to help them remember and understand learning content more effectively. However, they also share concerns about the cost of these devices and the reliance on sighted people, as sighted individuals cannot use these devices to help them in inevitable situations.

In addition, regarding inaccessible content, it is a challenge to promote accessible literacy to every learning resource producer. For instance, BLV students believe that for educational videos aimed at self-directed learning from BLV students, as P4 stated, \textit{``...Even if some of the material in this video cannot be read by a screen reader, some alternatives should be available, such as supporting text...''}.

\section{Discussion}
In this paper, we conducted one formative study and a two-part qualitative study to comprehensively investigate higher learning based on accessible technology in Chinese `Blind Colleges'. We emphasize the importance of accessible technology in this context and reveal a variety of practical difficulties and students' perceptions concerning current non-inclusive and future inclusive higher education supported by accessible technologies. We reflect on our findings and discuss the implications for future interventions involving accessible technology in higher BLV education in undeveloped and developing areas.
\subsection{Seeking Equality Learning in Non-Inclusive `Blind Colleges'}
In the findings section, we have illustrated the reasons, implementation, preferences, and challenges of BLV students using accessible technologies for studying in a `Blind Colleges' environment. Overall, we found that current accessible technologies could address most of the general learning needs of BLV students, such as attending lectures, writing assignments, reading files, and taking notes. Our participants also expressed a significant need for learning opportunities based on their interests beyond their major, which also depends on accessible technologies, educational software, and resources. This desire for embracing inclusive education prompts us to rethink the equality of `Blind Colleges'.

First, there is the issue of inequality in educational opportunities, where BLV individuals do not have a free and equal right to choose their specializations in these developing regions' `Blind Colleges'. They are eager to learn more specialties, but they are also concerned about their lack of competitiveness. Accessible technology has provided a bridge to equality for BLV students in this process, but the bridge is incomplete. Much of the current software and resources do not consider the learning needs of accessible students. As our participants suggested, the creators of these resources often do not consider that BLV students might wish to study a particular major. In other words, BLV students in `Integrated Colleges' can easily resolve their difficulties because they can easily seek help from sighted peers. However, this becomes inaccessible for BLV students in `Blind Colleges'. This inequality in educational opportunities is compounded by societal attitudes and technological imperfections, necessitating that BLV students exert more effort to access equal learning opportunities as those granted to them by right.

Second, there are issues of unequal access to education. With the importance of accessible technology in `Blind Colleges', BLV students are compelled to use accessible technology for learning. However, BLV students come from diverse backgrounds, and there are gaps in their ability to operate accessible technology. For those who are not proficient, this disparity creates a sense of embarrassment about their own learning when compared to peers who are adept. Such technology creates pressure for them to learn at an unequal pace, even in an educational environment meant to be equal for the BLV community. Additionally, higher education is inherently more challenging than basic education, and the need for BLV people to adapt to both the accessible technology and the complex knowledge they need to learn further exacerbates the inequality of learning. Thus, we advocate that all higher education institutions should pay more attention to teaching BLV students with accessible technology to reduce the bias from their unique pathway.

In summary, we believe that in such an educationally unequal environment, mastering accessible technology and acquiring knowledge become equally important for BLV students who seek to access equal learning opportunities.

\subsection{Transiting from Non-Inclusive `Blind Colleges' to Inclusive Higher Education}

As our research has found, the advancement of accessible technology does present a significant and sustained advancement of higher education for the BLV individuals in terms of effectiveness and usability. However, how does the transition from a non-inclusive `Blind Colleges' to an inclusive, integrated school advance in practice, given the influence of different social, human rights, and ecological factors under the diversity of the BLV community, higher education, and developing regions? We discuss the following three insights later this section.

First, a significant portion of non-inclusive higher education for BLV people is influenced by non-inclusive primary education. In our research case, we found that BLV students face strong difficulties with learning abilities, access to technology, and school choice. Because they were not exposed to inclusive education at an early age and because teachers often presented content at a simpler level, these students were unable to pursue advanced inclusive education even if they wanted to. We believe that higher BLV education cannot be discussed and governed in isolation; basic education needs to be inclusive, with more counseling, information synchronization, and social support provided to BLV students to help them learn about higher education in advance of their transition. Even if integrated schools are not feasible in some regions due to economic and social reasons, the practical use of accessible technology should be incorporated into the education of BLV people earlier, beyond simply teaching how to use computers and Braille displays.

Second, we concur with previous literature that the promotion of inclusive education relies not only on advances in accessible technology but also on the concerted efforts of policy decision-makers \cite{hewett2020Balancing, forlin2013changing}. Our research has found that social and policy factors more decisively influence equal higher education for the BLV than technological updates. In China, for example, our study found that even though the Chinese government allows BLV people to take the gaokao, there are few integrated schools available for BLV people to enroll in. Thus, in a context where BLV people generally take school-organized exams, teachers see no need for students to take the gaokao, creating a barrier. We believe that it is a higher priority, from a societal perspective, for more people to know and understand the abilities, aspirations, and wishes of BLV people. Our study also illustrates the need for accessible technology interventions to be designed and deployed in a way that fully considers the actual learning and acceptance abilities of BLV students, as well as the social and cultural environments in which they live, rather than merely replicating policies.

Lastly, higher BLV education should be brought more in line with the education of the sighted. Since the first `Blind School,' the Perkins School for the Blind, began \cite{french2004perkins}, there has been an ongoing conversation about the environment in which BLV people should learn and what specialties they should study. Accessible technological interventions have certainly provided BLV people with the ability to access more specialties, such as computer science \cite{Stefik2019coding, Lininger2008ComputersCS}. However, our research has found that BLV people have more individualized and interest-driven learning expectations, and they are assisted by technology. Likewise, they envision a greater variety of professional and vocational choices for themselves. Building on previous literature on the importance of maintaining special education schools for Blind in some regions \cite{miyauchi2022Keeping}, we propose that in developing regions, higher `Blind Colleges' should not be abandoned in the short term, but rather that more educational choices and pathways should be made available, drawing on the diversity of careers for the BLV and the mainstream of careers for the sighted in the regional community.

\subsection{Design Implications For BLV Accessible Educational Technologies}
It is worth noting that when we compare our results with previous studies, many of the same issues persist regarding accessible technologies, even though a decade has passed. We aligned our findings with the worldwide problems identified by previous researchers, such as the inconsistent experience of visibility and invisibility (i.e., accessible PDF, notes, coursework, and examinations), difficulties in comprehending classroom instructions delivered through computers, and concerns about potential judgment from sighted individuals regarding their computer work \cite{morina2017inclusive, Sue2018University, fichten2009accessibility, lourens2016s}. Thus, based on our findings and the discussion above, we propose a series of design implications for accessible educational technologies to enhance their practical efficacy and effectiveness, especially in BLV higher education.

\begin{itemize}
    \item \textbf{Facilitating Access to Educational Materials for BLV Students:} Based on feedback from our participants, the limited availability of accessible resources leads to cumbersome learning processes. This indicates the need for improvement on two fronts. Firstly, we suggest establishing an accessible resource platform where individuals can upload BLV-friendly electronic learning materials, providing a shared channel for resource acquisition. Secondly, we recommend major video platforms to introduce accessibility sections or label BLV-friendly videos (e.g., "This video contains alternative text"), making it easier for BLV individuals to identify valuable resources. These measures aim to alleviate the challenges faced by BLV students in accessing educational materials and enhance their learning experiences.

    \item \textbf{Enhancing Information Processing Efficiency through Human-AI Collaboration:} Due to we discovered the low-efficiency learning process, we suggested using more collaborative Human-AI approaches to support BLV students in their learning efficiency \cite{Amershi2019HumanAi, Wang2020humanAi}. We provided several scenarios in which BLV students can benefit from AI technologies. Such as 1) Using AI to extract key information for note organization during the course review; 2) Using AI to extract keywords and summarize abstracts from the textual information, breaking the linear reading format and thus increasing reading efficiency; 3) Using AI-Generated Content (AIGC) software \cite{wu2023aigenerated}, such as text to images \cite{pm2023review, frolovAdversarialTexttoimageSynthesis2021}, video \cite{wang2023modelscope}, audio \cite{Campos2023audio}, tables and slides \cite{Spataro2023Copilot} to empower BLV students in their content creation needs and facilitate interaction with sighted individuals. Overall, the field of designing BLV-AI collaboration systems is an emerging topic with many uncovered challenges and potential benefits in efficiency improvement \cite{Lee2022RSA}. Even though researchers have already proposed some AIGC tools for the BLV community \cite{Huh2023GenAssist}, how these tools meet BLV students' higher education needs still needs to be investigated.

    \item \textbf{Meeting the Need by Flexibility Methods:} As our findings illustrate, BLV students often struggle with accessibility software that is not usable or recognizable, thus hindering their learning. We suggest future accessibility technologies incorporate more comprehensive approaches, such as screen reading software with integrated OCR (Optical Character Recognition) for image recognition, or screen sharing and remote control features that make it easier to request help from a sighted person. Additionally, according to the section \ref{Non-Specialized Vision of Accessible Technologies} which discusses BLV students' preference for non-specialized software, we also recommend that future accessible system developers focus on supporting accessibility features within the original software itself. This includes designing user interfaces and shortcuts that are universally accessible and also creating customized opportunities for visually impaired users. In other words, BLV students would prefer using software that is accessible to the general public, which would be more conducive to their inclusion in society \cite{morina2017inclusive, forlin2013changing}.
\end{itemize}

\subsection{Limitation and Future Work}
However, we acknowledge several limitations within our study. Firstly, the homogeneity of the participants' majors, with only music and acupuncture and tuina science being available to BLV students in Chinese `Blind Colleges', raises concerns about the diversity of our findings. In other developing countries and underdeveloped areas, students may have different educational backgrounds and technical literacy, which create unique challenges and perceptions because of local cultural and educational factors.
Additionally, the study was conducted within a `Blind Colleges' with an extremely low acceptance rate (less than 10\%), indicating that the academic abilities of our participants may be above average for BLV students. This potential bias toward a higher academic level within our sample could affect the applicability of our findings to BLV students in different educational settings.
We suggest that future research should consider including BLV students from various areas and educational contexts. Systematicly evaluate the consistency, reliability, and ethics of accessible use for learning by BLV students in `Blind Colleges' setting to promote future inclusive education in developing and underdeveloped areas.
In conclusion, our work is akin to making an essential addition to the literature on students from developing and undeveloped areas by capturing the state of opinions of Chinese students rather than representing all BLV higher education students.

\section{Conclusion}
In this paper, we conducted a two-part empirical study to comprehensively explore the practices, perceptions, and challenges of using accessible technologies for studying in the context of Chinese `Blind Colleges'. Our findings highlight the important roles of accessible technologies in `Blind Colleges', yet also reveal multiple practical challenges due to inaccessibility. We also uncovered a dilemma regarding the vision of future inclusive learning, influenced by imperfect accessible technologies and the impact of long-term non-inclusive learning experiences. Drawing on these findings, we discussed the role of accessible technologies in transitioning `Blind Colleges' into `Integrated Colleges' for future BLV higher education in developing areas and provided several design considerations for educational accessible technologies. This work contributes to the HCI communities by raising awareness of the unbalanced educational technology situation in undeveloped and developing areas, emphasizing the importance of advancing technological and educational innovation.

\begin{acks}
This work was supported by the Foundation of the Ministry of Education of China under Grant number 23YJCZH092.
\end{acks}

\section*{Disclosure Statement}
This work does not have any conflicts of interest to disclose.

\section*{Generative Artificial Intelligence (AI) Disclosure}
As non-native English speakers and researchers, we kindly use artificial intelligence (ChatGPT-4.0) to only check grammar (but not rewrite) for sentence or word accuracy. We did not use AI to directly generate any of the content in this article.
\bibliographystyle{ACM-Reference-Format}
\bibliography{Main}

\appendix
\section{The Classwork of Mock-up Course Workshop}
\label{Classwork}
The HCI professor delivered a 30-minute online lecture, delving into incorporating the haptic into music. In the discourse, the professor introduced how to use interdisciplinary design that combined accessibility and HCI to better benefit inclusive life. The details of the tasks can be found in Table 3.
\begin{table}[h]
\centering
\label{Table3}
\caption{Tasks, Instructions and Area of focus about the workshop assignments}
\begin{booktabs}{
  column{1}={c},
  colspec={c X[l] l},
  cells={m},
  width=\linewidth,
  hspan=minimal
}
\toprule
\#Tasks & Instructions                                                                                                                                                                                                           & Area of focus      \\
\midrule
1      & Summarize a method to help hearing-impaired students understand the rhythm of `segmentation' through haptic design.                                                                                                             & Summarization      \\
\midrule
2      & Using the internet, read and summarize relevant literature on `haptic' and `music perception'.                                                                                                                         & Information Search \\
\midrule
3      & {Discuss how to apply the discovery of “tactile channel perception of sound” to other scenarios. Utilizing news, literature, research reports and other data sources to support the group’s finding and conclusion.} & Collaboration   \\
\bottomrule
\end{booktabs}
\end{table}

\section{Selection and Participation of BLV Students}
This study involving participants with blind or low vision followed rigorous ethical guidelines and comprehensive informed consent procedures. The recruitment of the participants was facilitated through our lab's collaboration with a Chinese `Blind College.' The researchers provided the study's objective and procedures, ensuring the prospective participants were well-informed before filling out a demographic survey and providing their informed consent. Throughout the study, researchers ensured that the data collected from the participants would be anonymous and for study purposes only. To create a safe and comfortable environment for the participants, the study was conducted within the familiar setting of the students' university classrooms. IRB approved the entire procedure.

\section{Interview Scripts}
\label{Interview}

\textbf{Introduction:}

Welcome! Thank you for participating in this interview session. We greatly appreciate your time and insights. Today, we would like to discuss your daily learning habits, preferences for electronic devices, and thoughts on assistive technology for blind and low vision individuals.

\textbf{Section 1: Daily Learning Habits and Challenges}
\begin{enumerate}
\item Can you describe your typical daily routine involving learning and other activities?
\item What subjects or courses do you usually study?
    \begin{itemize}
    \item Which courses do you find particularly challenging, and why?
    \end{itemize}
\item Could you share your typical study habits?
    \begin{itemize}
    \item How do you usually approach learning in a classroom setting?
    \item Do you engage in activities like review, homework, or preparation after returning home? Are there any inconveniences compared to studying at school?
    \end{itemize}
\end{enumerate}

\textbf{Section 2: Usage Patterns and Preferences for Accessible Technologies}
\begin{enumerate}
\item How do you incorporate accessible technologies into your daily life?
    \begin{itemize}
    \item What are your habits and preferences regarding their usage? (If possible, could you demonstrate?)
    \end{itemize}
\item How do you utilize existing assistive learning electronic products or applications?
    \begin{itemize}
    \item What tasks do you perform with them?
    \item How would you describe your user experience? Are there any areas of difficulty or challenges encountered?
    \end{itemize}
\end{enumerate}

\textbf{Section 3: Awareness, Needs, and Expectations Regarding Blind-Friendly accessible technologies}
\begin{enumerate}
\item What is your familiarity with blind-friendly accessible technologies or similar assistive learning products?
    \begin{itemize}
    \item Have you had experience with any comparable assistive learning or daily living products?
    \item How do you envision a blind-friendly accessible technologies? (Show a sample if available) Does it meet your expectations?
    \end{itemize}
\item What scenarios or situations do you believe necessitate the use of accessible technologies?
    \begin{itemize}
    \item How would you utilize accessible technologies? (Consider required auxiliary devices, support personnel, and teaching methods)
    \end{itemize}
\item What are your expectations for accessible technologies?
    \begin{itemize}
    \item In what scenarios or situations do you believe such a device would assist you in your daily life? What aspects of your life would it benefit?
    \end{itemize}
\end{enumerate}

\textbf{Conclusion:}

Thank you for sharing your valuable insights with us today. Your input will be instrumental in developing technologies that cater to the needs of blind and low vision individuals. If you have any additional comments or suggestions, please feel free to share them. We truly appreciate your participation.

\end{document}